\documentclass[a4paper]{jpconf}
\usepackage{graphicx}

\usepackage{amssymb}
\usepackage{amsmath}
\usepackage{graphicx}
\usepackage[normalem]{ulem}
\usepackage[dvips]{color}
\renewcommand\sout{\bgroup \color{red} \ULdepth=-.5ex \ULset}

\begin{document}
\title{Probing Nuclear Symmetry Energy and its Imprints on Properties of Nuclei, Nuclear Reactions, Neutron Stars and Gravitational Waves}

\author{Bao-An Li$^1$, Lie-Wen Chen$^{1,2}$, Farrukh J. Fattoyev$^1$,\\ William G. Newton$^1$ and Chang Xu$^{1,3}$}
\address{$^1$Department of Physics and Astronomy, Texas A$\&$M
University-Commerce, Commerce, Texas 75429-3011, USA}
\address{$^2$Department of Physics, Shanghai Jiao Tong University,
Shanghai 200240, China}
\address{$^3$Department of Physics, Nanjing University, Nanjing 210008, China}

\ead{Bao-An.Li@Tamuc.edu}

\begin{abstract}
Significant progress has been made in recent years in constraining
nuclear symmetry energy at and below the saturation density of
nuclear matter using data from both terrestrial nuclear experiments
and astrophysical observations. However, many interesting questions
remain to be studied especially at supra-saturation densities. In
this lecture note, after a brief summary of the currently available
constraints on nuclear symmetry energy near the saturation density we
first discuss the relationship between the symmetry energy and the
isopin and momentum dependence of the single-nucleon potential in
isospin-asymmetric nuclear medium. We then discuss several open issues regarding
effects of the tensor force induced neutron-proton short-range correlation (SRC)
on nuclear symmetry energy. Finally, as an example of the
impacts of nuclear symmetry energy on properties of neutron stars
and gravitational waves, we illustrate effects of the high-density
symmetry energy on the tidal polarizability of neutron stars in
coalescing binaries.
\end{abstract}

\begin{figure}[htb]
\includegraphics[scale=1.3]{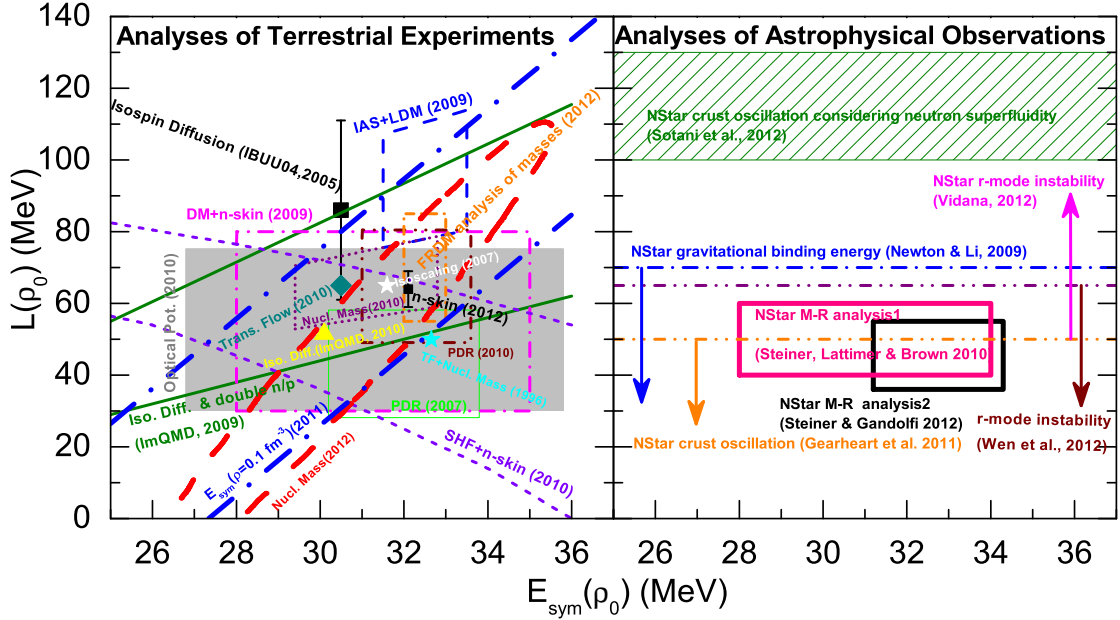}\label{Esym&L}
\caption{Density slope versus the magnitude of the symmetry energy
at saturation density extracted from analyzing terrestrial
experiments (left window) and astrophysical observations (right
window). The analyses of terrestrial experiments include (1)
analyses of isospin diffusion experiments with $^{124}$Sn+$^{112}$Sn
at 50 MeV/A within the Isospin-Dependent Boltzmann-Uehling-Ulenbeck
(IBUU04-2005)\cite{chen05a,li05a}, (2) the isospin diffusion and
neutron/proton ratio of pre-equilibrium nucleon emissions in
$^{124}$Sn+$^{112}$Sn reactions at 50 MeV/A within the Improved
Molecular Dynamics (ImQMD-2009) model \cite{tsa1,tsa2,Fami}, (3)
$^{124}$Sn+$^{112}$Sn reactions at 35 MeV/A within the Improved
Molecular Dynamics (ImQMD-2010) model \cite{sun10}, (4) isoscaling
(isoscaling-2007) \cite{shet}, (5) energy shift of isobaric analogue
states within liquid drop model (IAS+LDM-2009)\cite{Pawel09}, (6)
neutron-skins of several heavy nuclei using the droplet model
(DM+n-skin (2009))\cite{Cen09,War09}, or the Skyrme-Hartree-Fock
(SHF+n-skin) approach \cite{Chen10}, or the phenomenological
approach (n-skin 2012) \cite{agr12}, (7) pygmy dipole resonances
(PDR 2007 and 2010) in $^{209}$Pb, $^{68}$Ni and $^{132}$Sn
\cite{Kli07,Car10}, (8) the nucleon global optical potentials
(Optical Pot. 2010)~\cite{XuLiChen10a}, (9) atomic masses analyzed
by Myers and Swiatecki using the Thomas-Fermi model (TF+Nucl. Mass
(1996) \cite{mye96}, (10) atomic masses analyzed by M\"oller et al.
using the finite-range droplet model (FRDM) \cite{Moller}, (11)
atomic masses analyzed by Liu et al. (Nucl. Mass (2010)
\cite{Mliu10}, (12) atomic masses analyzed by Lattimer and Lim
(Nucl. Mass (2012) \cite{Jim}, (13) Anti-symmetrized Molecular
Dynamics (AMD) analyses of transverse flow of inter mediate mass
fragments (Trans. Flow (2010)) \cite{Koh10}, (14) empirical value of
the symmetry energy at $\rho=0.1 \, {\rm fm}^{-3}$ ($E_{\rm
sym}(\rho=0.1 \, {\rm fm}^{-3}) (2011)$ \cite{LWChen11}. The
analyses of astrophysical observations include (15) the mass-radius
correlation of neutron stars (NStar analysis1 and analysis2)
\cite{Steiner10,Steiner12}, (16) gravitational binding energy of
neutron stars (Newton \& Li, 2009) \cite{Newton09}, (17) torsional
oscillations of neutron star crust analyzed by Gearheart et al.
\cite{Mike} and Sotani et al. \cite{Sot12}, (18) the r-mode
instability of neutron stars analyzed by Wen et al. \cite{Wen12} and
Vidana \cite{Vid12}. Similar plots from selecting different sets of
constraints available in the literature at the time can be found in
Refs.~\cite{XuLiChen10a,Tsang12,Lattimer12,Chen12}.}
\end{figure}
\section{Introduction}
The Equation of State (EOS) of neutron-rich nucleonic matter can be
written within the parabolic approximation in terms of the binding
energy per nucleon at density $\rho$ as
\begin{equation}
E(\rho ,\delta)=E(\rho ,\delta =0)+E_{\rm sym}(\rho )\delta
^{2}+\mathcal{O}(\delta^4) \ ,
\end{equation}
where $\delta\equiv(\rho_{n}-\rho _{p})/(\rho _{p}+\rho _{n})$ is
the neutron-proton asymmetry and $E_{\rm sym}(\rho)$ is the
density-dependent nuclear symmetry energy. The latter is among the
most uncertain properties of dense, neutron-rich nucleonic matter.
It has important ramifications for many interesting questions in
both astrophysics and nuclear physics. Thanks to the hard work of
many people in both astrophysics and nuclear physics communities,
significant progress has been made in recent years in constraining
the density dependence of nuclear symmetry energy using data from
both terrestrial nuclear experiments and astrophysical observations.
A deeper understanding about the underlying physics governing the
density dependence of nuclear symmetry energy was also obtained from
various theoretical studies. However, many challenging questions
remain to be answered. As an illustration of our current
understanding about nuclear symmetry energy near saturation density
$\rho_0$, shown in Fig. 1 are the available constraints on the slope
$L(\rho_0) \equiv 3 \rho \frac{\partial E_{\rm sym}(\rho)}{\partial
\rho}\big|_{\rho_0}$ versus symmetry energy $E_{\rm sym}(\rho_0)$ at
$\rho_0$ from analyses of both terrestrial nuclear experiments and
astrophysical observations. Besides experimental error bars, there
are some model dependences in most analyses and not all model
assumptions are equally valid. Even assuming all published results
are equally physical, given the still widely scattered constraints
it is difficult to calculate a community-average of the $E_{\rm
sym}(\rho_0)$ and $L(\rho_0)$ with physically meaningful error bars
at this time. With all due respects to conclusions others may have
drawn, in our obviously biased opinion, $E_{\rm sym}(\rho_0) = 31
\pm 2$ MeV and $L(\rho_0)=50 \pm 20$ MeV are probably the best
empirical values with the optimistic error bars we can currently
conclude. In the following, using three examples we illustrate some
recent progress in understanding the underlying physics governing
the density dependence of nuclear symmetry energy, why the symmetry
energy is still very uncertain, and how to probe the high density
behavior of nuclear symmetry energy. The materials presented here
are mostly taken from our recent work published originally in
Refs.~\cite{XuLiChen10a,Mike,Xu-tensor,xuli2,Rchen,tensor12,Will1,Will2,Farooh1,Farooh2,LCK08}.

\section{The relationship between nuclear symmetry energy and single-nucleon mean-field potential}
What is the direct relationship between the symmetry energy and the
isoscalar and isovector parts of the single-nucleon potential
$U_{\rm n/ \rm p}(\rho,\delta,k)$? An answer to this question helps
us better understand why the symmetry energy is still very
uncertain. To our best knowledge, this question was first studied by
Brueckner, Dabrowski and Haensel \cite{bru64,Dab72,Dab73,Dab74}
using K-matrices within the Brueckner theory in the 1960's. More
recently, it was studied by Xu et al. \cite{xuli2} and Chen et al.
\cite{Rchen} using the Hugenholtz-Van Hove (HVH) theorem \cite{hug}.
This is an important question for several reasons. First of all, in
both nuclear physics and astrophysics, one of the ultimate goals is
to understand the isospin dependence of strong interaction. Both the
symmetry energy and the single-particle potential are determined by
the same underlying strong interaction. Their relationship can thus
help us better understand why the symmetry energy is still very
uncertain, and connect with the QCD theory of nuclear strong
interaction. Theoretically, one usually derives both the
single-nucleon potential and the symmetry energy from a model energy
density functional constrained by empirical properties of nuclear
matter and finite nuclei. However, the single-nucleon potential is
often the one directly tested by comparing model calculations with
experimental data. For example, the single-particle potential is the
input for shell model calculations of nuclear structure and
transport model simulations of nuclear reactions. Therefore, being
able to know directly the corresponding symmetry energy from the
single-particle potential without first going through the procedure
of constructing the energy density functional is advantageous. For
example, from nucleon-nucleus scattering and (p, n) charge exchange
experiments one can directly extract from the data both the
isoscalar and isovector nucleon optical potentials at normal
density. One can then easily calculate the symmetry energy and its
density slope at normal density directly from the optical potentials
as demonstrated recently in Ref. \cite{XuLiChen10a}.

According to the well-known Lane potential \cite{Lan62}, the
neutron/proton single-particle potential $U_{n/p}(\rho,k,\delta)$
can be well approximated by
\begin{equation}
U_{\rm n/ \rm p}(\rho,k,\delta)\approx U_0(\rho,k) \pm U_{\rm
sym}(\rho,k)\delta \ ,
\end{equation}
where the $U_0(\rho,k)$ and $U_{\rm sym}(\rho,k)$ are, respectively,
the isoscalar and isovector (symmetry) potentials for nucleons with
momentum $k$ in nuclear matter of isospin asymmetry $\delta$ at
density $\rho$. It has been shown that the nuclear symmetry energy
can be explicitly expressed as
\cite{XuLiChen10a,xuli2,Rchen,bru64,Dab72,Dab73,Dab74}
\begin{equation}
E_{\rm sym}(\rho) = \frac{1}{6} \frac{\partial(t+U_0)}{\partial
k}\bigg|_{k_{\rm F}}\cdot k_{\rm F} + \frac{1}{2}U_{\rm
sym}(\rho,k_{\rm F}) \ ,
\end{equation}
where $t(k)=\hbar ^2 k^2 / 2m$ is the kinetic energy and $k_{\rm
F}=(3\pi^2\rho/2)^{1/3}$ is the nucleon Fermi momentum in symmetric
nuclear matter at density $\rho$. The slope of nuclear symmetry
energy at an arbitrary density $\rho$ can be written as
\cite{XuLiChen10a,xuli2,Rchen}
\begin{eqnarray}
\nonumber L(\rho) &\equiv& 3 \rho \frac{\partial E_{\rm
sym}(\rho)}{\partial \rho}\bigg|_{\rho} = \\ \ &=& \frac{1}{6}
\frac{\partial (t+U_0)}{\partial k}\bigg|_{k_{\rm F}} \cdot k_{\rm
F} + \frac{1}{6} \frac{\partial^2 (t+U_0)}{\partial
k^2}\bigg|_{k_{\rm F}} \cdot  k_{\rm F}^2 + \frac{3}{2} U_{\rm
sym}(\rho,k_{\rm F})+ \frac{\partial U_{\rm sym}}{\partial
k}\bigg|_{k_{\rm F}} \cdot k_{\rm F} \ .
\end{eqnarray}
Moreover, the relative neutron-proton effective mass is
\begin{eqnarray}
\frac{m_{\rm n}^*-m_{\rm p}^*}{m}&=&-2 \delta \frac{m}{\hbar^2
k_{\rm F}} \frac{dU_{\rm sym}}{dk}\bigg|_{k_{\rm F}} \bigg/ \left[1+
2 \frac{m}{\hbar^2 k_{\rm F}}
\frac{dU_0}{dk}\bigg|_{k_{\rm F}} \right] \nonumber \\
&=&-2 \delta \frac{m}{\hbar^2  k_{\rm F}}\frac{dU_{\rm
sym}}{dk}\bigg|_{k_{\rm F}} \bigg/ \left[1+ 2
\left(\frac{m}{m_0^*}-1\right) \right] \label{npemass} \ .
\end{eqnarray}
We emphasize that this relationship is valid only at the mean-field
level. Taking into account the tensor force induced neutron-proton
short range correlation, the kinetic part of the symmetry energy
might be reduced significantly from the Fermi gas model prediction
as we shall discuss in detail in the next section. The above
expressions for $E_{\rm sym}(\rho)$ and $L(\rho)$ in terms of the
isoscalar and isovector single-particle potentials are particularly
useful for extracting the symmetry energy and its density slope from
terrestrial nuclear laboratory experiments. While the density and
momentum dependence of the isoscalar potential $U_{0}(\rho,k)$ has
been relatively well determined up to about 4 to 5 times the normal
nuclear matter density $\rho_0$ using nucleon global optical
potentials from nucleon-nucleus scatterrings as well as kaon
production and nuclear collective flow in relativistic heavy-ion
collisions \cite{dan02,Fuc06Kaon}, our current knowledge about the
symmetry potential $U_{\rm sym}(\rho,k)$ is rather poor especially
at high density and/or momenta
\cite{LCK08,Fuchs05,zuo05,ria05,ria06,Gio10}. Experimentally, there
is some constraints on the symmetry potential only at normal density
for low energy nucleons up to about 100 MeV obtained from
nucleon-nucleus and (p, n) charge exchange reactions
\cite{XuLiChen10a}.
\begin{figure}[htb]
\begin{center}
\begin{minipage}{18pc}\label{GOP}
\includegraphics[scale=0.7]{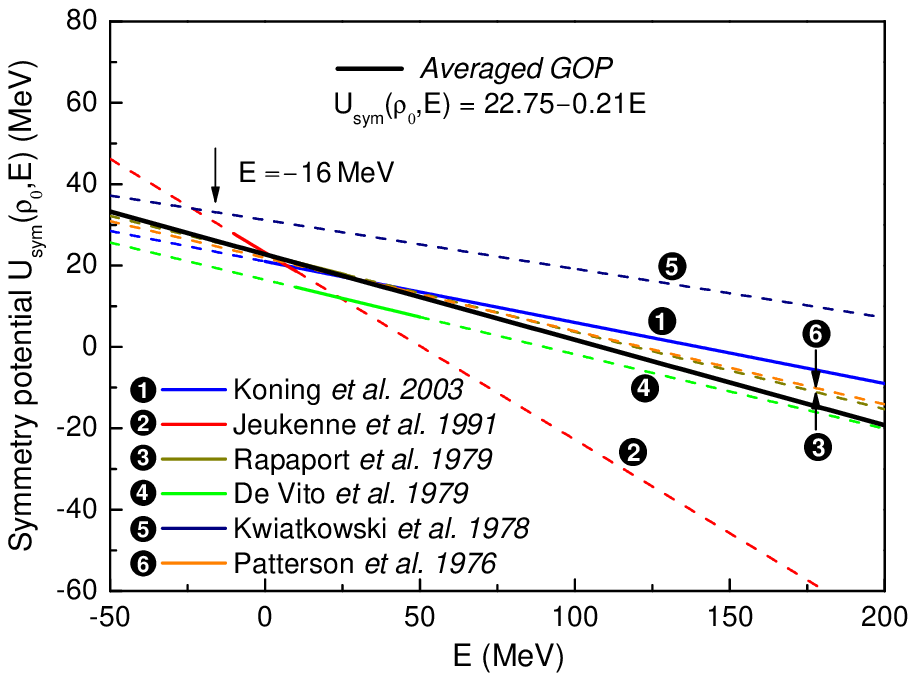}
\end{minipage}
\begin{minipage}{14pc}
\includegraphics[scale=0.7]{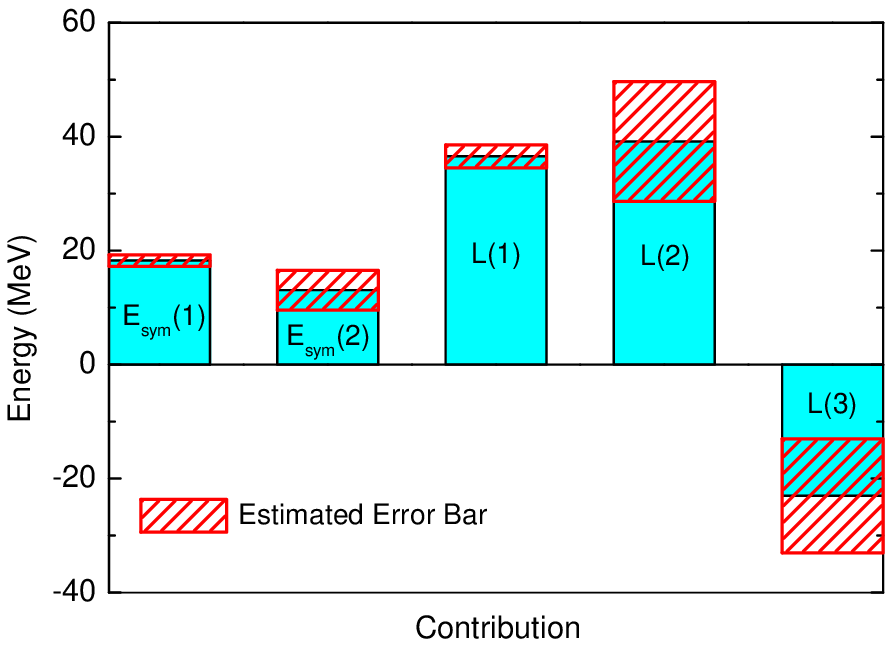}
\end{minipage}
\caption{\label{Xu1} {\protect\small \textbf{Left Window}: Energy
dependence of the nuclear symmetry potential $U_{\rm sym}(\rho_0,E)$
at normal density from different global optical model fits. The
solid lines are in the energy ranges where the original analyses
were made while the dashed parts are extrapolations. \textbf{Right
Window}: The magnitude of each term in the nuclear symmetry energy
$E_{\rm sym}(\rho_0)$ and its density slope $L(\rho_0)$ at normal
nuclear density.} Taken from Ref. \cite{XuLiChen10a}.}
\end{center}
\end{figure}
Shown in the left window of Fig. \ref{GOP} are all the energy dependent
symmetry potentials in the literature \cite{XuLiChen10a}. Assuming
that these various global energy dependent symmetry potentials are
equally accurate and all have the same predicting power beyond the
original energy ranges in which they were analyzed, an averaged
symmetry potential of
\begin{equation}
U_{\rm sym}(\rho_0,E)=22.75-0.21E\label{Ubest}
\end{equation}
was obtained (thick solid line in the left window of Fig.
\ref{GOP}). It represents the best fit to the global symmetry
potentials constrained by the world data up to date. With this best
estimate for the $U_{\rm sym}(\rho_0,E)$, Xu et al. found that
$E_{\rm sym}(\rho_0)= 31.3 \pm 4.5$ MeV and $L(\rho_0)=52.7 \pm 22.5
$ MeV. Shown in the right window of Fig. 2 are the various
contributions to the $E_{\rm sym}(\rho_0)$ and $L(\rho_0)$. The
$E_{\rm sym}(1)=\frac{1}{3} \frac{\hbar^2 k_{\rm F}^2}{2 m_0^*}$
denotes the kinetic energy term with the effective mass $m_0^*$ and
the $E_{\rm sym}(2)=\frac{1}{2} U_{\rm sym}(\rho_0, k_{\rm F})$ is
the symmetry potential contribution. It is seen that the two terms
are comparable. Their respective uncertainties are marked by the red
boxes. The $L(\rho_0)$ has three terms: $L(1)=\frac{2}{3}
\frac{\hbar^2 k_{\rm F}^2}{2 m_0^*}$, $L(2)=\frac{3}{2} U_{\rm
sym}(\rho_0,k_{\rm F}) $ and $L(3)= \frac{\partial U_{\rm
sym}(\rho,k)}{\partial k}\big|_{k_{\rm F}} k_{\rm F}$. While both
the $L(1)$ and $L(2)$ are positive, the $L(3)$ is negative because
of the decreasing symmetry potential with increasing energy. To our
best knowledge, extracting the symmetry energy and its density slope
directly from the optical potential is probably the most straight
forward approach available in the literature. However, the major
challenge of using this approach is our poor knowledge about the
momentum dependence of the isovector potential. From the symmetry
potential given in Eq. (\ref{Ubest}), Xu et al. extracted a
neutron-proton effective mass splitting of $(m_{\rm n}^*-m_{\rm
p}^*)/m=(0.32 \pm 0.15)\delta$. Interestingly, from the dispersive
optical model anylysis of some new data on neutron-nucleus
scattering, R. Charity recently also found a  value of $(m_{\rm
n}^*-m_{\rm p}^*)/m=0.32 \delta$ with an error bar to be determined
\cite{Bob}. The extracted constraint on $E_{\rm sym}(\rho_0)$ versus
$L(\rho_0)$ from the global nucleon optical potentials are compared
with constraints extracted from other approaches \cite{Chen12} in
Fig.\ 1. It is consistent with the ones from most other approaches.

\section{Effects of tensor force induced neutron-proton short-range correlation on nuclear symmetry energy}
From Fig. 1 and the related references, it is clear that there are
still appreciable uncertainties about the density dependence of
nuclear symmetry energy even around the saturation density.
Moreover, at supra-saturation densities, even the tendency of the
symmetry energy remains controversial \cite{LCK08}. So, why is the
nuclear symmetry energy, especially at supra-saturation densities,
so uncertain? Of course, the answer itself is model dependent.
Generally, besides our poor knowledge about the isospin dependence
of strong interaction in dense neutron-rich medium, different
approaches used in treating nuclear many-body problems in various
models contribute to the divergence of the predicted symmetry energy
especially at supra-saturation densities. Nonetheless, there are
several key and commonly used physics ingredients that can affect
the predicted $E_{\rm sym}(\rho)$ in all theories. For instance, the
symmetry energy has a kinetic part. Often, it is assumed to be the
one predicted by the free Fermi gas model, i.e., $E_{\rm sym}^{\rm
kin}(\textrm{FG})(\rho)\equiv(2^{\frac{2}{3}}-1)(\frac{3}{5}
\frac{\hbar^2 k_{\rm F}^2}{2m})\approx 12.5(\rho/\rho_0)^{2/3}$.
Interestingly, it was first found recently within a phenomenological model
\cite{XuLiLi} that the tensor force induced high momentum tail in
the single-nucleon momentum distribution in symmetric nuclear matter
(SNM) reduces significantly the $E_{\rm sym}^{\rm kin}(\rho)$ to
values much smaller than the $E_{\rm sym}^{\rm
kin}(\textrm{FG})(\rho)$. In fact, the $E_{\rm sym}^{\rm kin}(\rho)$
can become zero or even negative if more than about $15\%$ nucleons
populate the high-momentum tail above the Fermi surface as indicated
by the recent experiments done at the Jefferson National Laboratory
(J-Lab) by the CLAS Collaboration \cite{CLAS}. This finding was recently
confirmed qualitatively by three independent studies using
the state-of-the-art microscopic many-body theories
\cite{vid,carb,lov}.
\begin{figure}[htb]
\centering
\includegraphics[width=15cm]{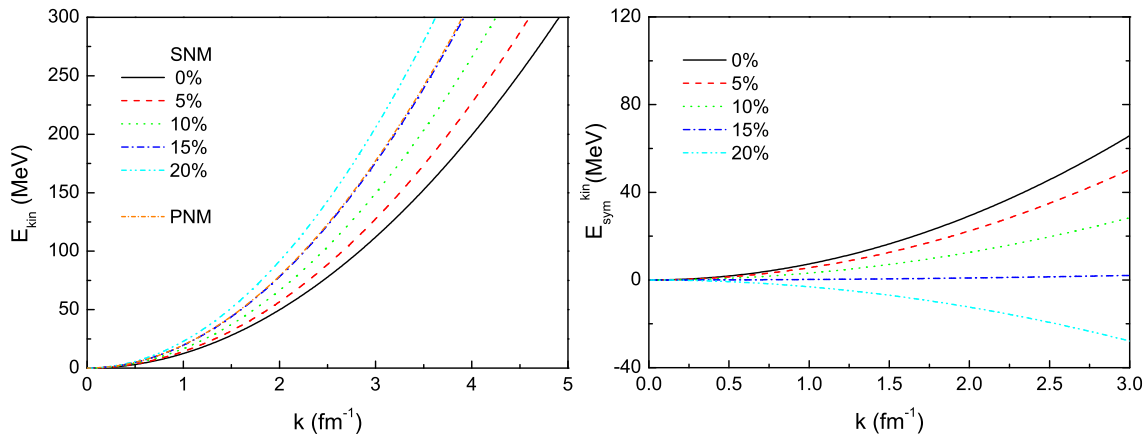}
\caption{\textbf{Left Window}: The average kinetic energy per
nucleon $E_{\rm kin}$ for pure neutron matter and symmetric nuclear
matter with different percentages of correlated nucleons
($\theta_{k>k_{\rm F}}$) as a function of Fermi momentum.
\textbf{Right Window}: The kinetic energy part of nuclear symmetry
energy with different percentages of correlated nucleons
($\theta_{k>k_{\rm F}}$) as a function of Fermi momentum. Taken from
Ref. \cite{XuLiLi}.} \label{Ekin}
\end{figure}
As discussed in detail by Xu et al. in Ref. \cite{XuLiLi}, the high
momentum tail in SNM increases the average kinetic energy of
nucleons to values above the free Fermi gas model prediction. While
in PNM (pure neutron matter), the Fermi gas prediction is a good
approximation. Shown in the left window of Fig. \ref{Ekin} is the
the average kinetic energy as a function of Fermi momentum for PNM
and SNM with the percentage of high momentum nucleons to be
$\theta_{k>k_{\rm F}}=0\%$, $5\%$, $10\%$, $15\%$, \textrm{and}
$20\%$, respectively. As one expects, the SRC increases the $E_{\rm
kin}$ significantly for SNM. More quantitatively, for SNM at the
saturation density corresponding to $k_{\rm F}=$1.33 fm$^{-1}$, the
$E_{\rm kin}$ with $\theta_{k>k_{\rm F}}=20\%$ ($E_{\rm kin}(k_{\rm
F})\simeq40$ MeV) is about twice of that ($E_{\rm kin}(k_{\rm
F})\simeq22$ MeV) for the free Fermi gas. However, the $E_{\rm kin}$
for PNM is the same as for the free Fermi gas. Consequently, the
tensor force induced high momentum tail in SNM affects the kinetic
part of the nuclear symmetry energy. In particular, if about $15\%$
nucleons in SNM are in the high momentum tail, it is seen that the
average kinetic energy is about the same in PNM and SNM. This leads
to an approximately zero kinetic symmetry energy as shown in the
right window of Fig. \ref{Ekin}. In many studies in both nuclear
physics and astrophysics, it is customary to write the total
symmetry energy as $E_{\rm sym}(\rho)=12.5(\rho/\rho_0)^{2/3}+E^{\rm
pot}_{\rm sym}(\rho)$ where the first term is the Fermi gas
prediction for the $E_{\rm sym}^{\rm kin}(\rho)$ and the $E^{\rm
pot}_{\rm sym}(\rho)$ is the potential contribution. In doing so,
however, one neglects completely effects of the tensor force on the
$E^{\rm kin}_{\rm sym}(\rho)$. For example, in transport model
analyses of heavy-ion reactions, the $E^{\rm pot}_{\rm sym}(\rho)$
is normally parameterized as a function of density. The
corresponding single-particle potential based on some energy density
functions is used as input to transport models. Thus, heavy-ion
reactions test directly the single nucleon potential. The kinetic
part of the symmetry energy based on the Fermi gas model prediction
is normally added by hand to the $E^{\rm pot}_{\rm sym}(\rho)$ in
fixing parameters in the EOS. The results shown in Fig.\ \ref{Ekin}
raise serious questions about this practice. Moreover, many
interesting questions regarding effects of the tensor-force induced
isospin dependence of short-range nucleon-nucleon correlations and
the uncertain short-range behavior of tensor force due to the $\rho$
meson exchange on the density dependence of nuclear symmetry energy
remain to be studied more systematically and self-consistently \cite{Xu-tensor,Lee11,Li11}.
\begin{figure}[htb]
\begin{minipage}{18pc}\label{LoveNo}
\includegraphics[scale=0.3]{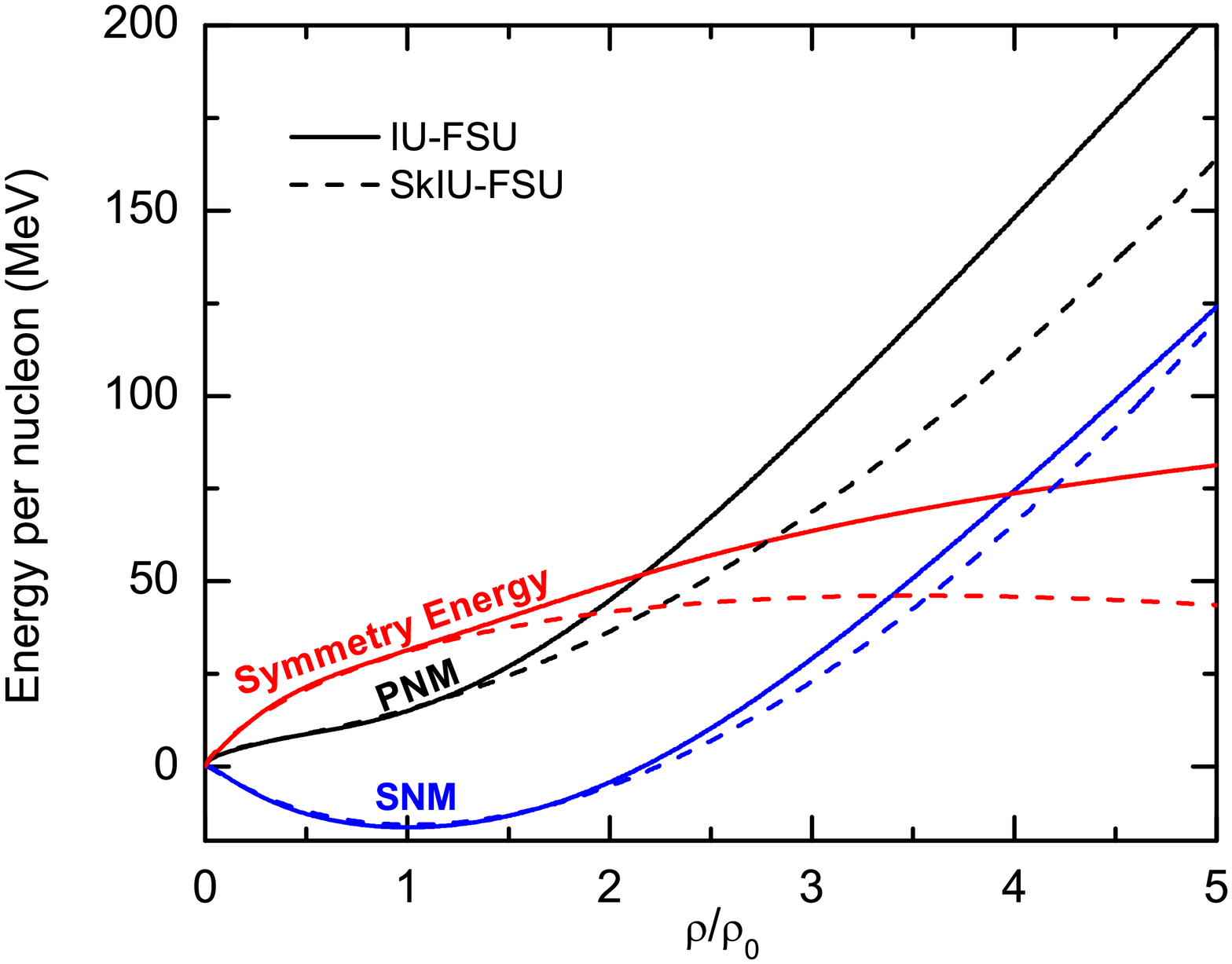}
\end{minipage}
\begin{minipage}{12pc}
\includegraphics[scale=0.3]{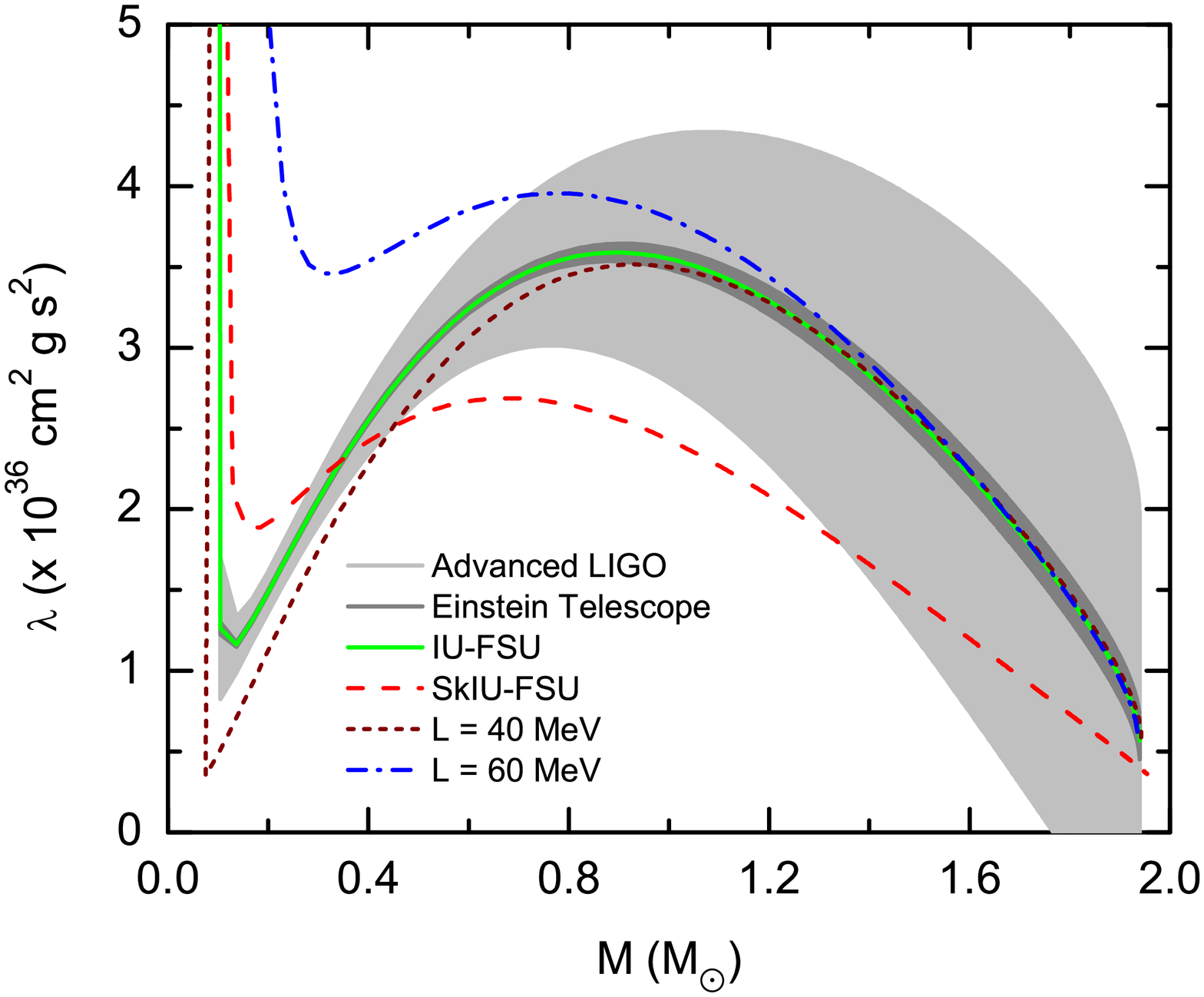}
\end{minipage}
\caption{{\protect\small \textbf{Left Window}: The EOS of symmetric
nuclear matter and pure neutron matter as well as the symmetry
energy as a function of density obtained within the IF-FSU RMF model
and the SHF approach using the SkIU-FSU parameter set. \textbf{Right
Window}: Tidal polarizability $\lambda$ of a single neutron star as
a function of neutron-star mass for a range of EOS that allow
various stiffness of symmetry energies. A crude estimate of
uncertainties in measuring $\lambda$ for equal mass binaries at a
distance of $D = 100$ Mpc is shown for the Advanced LIGO (shaded
light-grey area) and the Einstein Telescope (shaded dark-grey area)
}. Taken from Ref. \cite{Farooh2}.}
\end{figure}
\section{Probing the high-density symmetry energy with the tidal polarizability of neutron stars}
The high-density behavior of nuclear symmetry energy has long been
regarded as the most uncertain property of dense neutron-rich
nucleonic matter \cite{LCK08,Kut94,Kub99}. While several observables
have been proposed~\cite{LCK08} and some indications of the
high-density symmetry energy have been reported based on terrestrial
nuclear experiments~\cite{Xiao09,Paolo}, unfortunately, the
conclusions remain controversial. Interestingly, it was recently
proposed that the late time neutrino signal from a core collapse
supernova~\cite{Roberts} and the tidal polarizability \cite{Farooh2}
of canonical neutron stars in coalescing binaries are very sensitive
probes of the high-density behavior of nuclear symmetry energy.
Coalescing binary neutron stars are among the most promising sources
of gravitational waves (GW). One of the most important features of
the binary mergers is the tidal deformations of neutron stars, which
give us precious information about the neutron-star matter EOS
\cite{Flanagan:2008,Damour:2009,Damour:2010,Damour:2012,Hinderer:2008,
Hinderer:2010,Pannarale:2011,Postnikov:2010yn}. At the early stage
of an inspiral tidal effects may be effectively described through
{\sl the tidal polarizability} parameter
$\lambda$~\cite{Flanagan:2008, Damour:2009, Damour:2010,
Hinderer:2010} defined via $Q_{ij}= - \lambda \mathcal{E}_{ij}$,
where $Q_{ij}$ is the induced quadrupole moment of a star in binary,
and $\mathcal{E}_{ij}$ is the static external tidal field of the
companion star. As an example, shown in the left window of Fig. 4
are two models chosen to have the same EOSs for both SNM and PNM
using the IU-FSU RMF model and the SHF using the SkIU-FSU parameter
set~\cite{Farooh1}, i.e., they have the same symmetry energy at and
below the saturation density. At supra-saturation densities,
however, the symmetry energy with the IU-FSU RMF is significantly
more stiff above about $1.5\rho_0$. It is seen from the right window
of Fig. 4 that the two models predict significantly different
polarizability ($\lambda$) values in a broad mass range from 0.5 to
2 $M_{\odot}$. More quantitatively, for a canonical neutron star of
1.4 $M_{\odot}$, a $41.41\%$ change from $\lambda = 2.828\times
10^{36}$ (IU-FSU) to $\lambda= 1.657\times 10^{36}$ (SkIU-FSU) is
observed. As it was discussed in detail in Ref. \cite{Farooh2}, the
observed symmetry energy effect on the tidal polarizability is as
strong as its effect on the late time neutrino flux from the cooling
of proto-neutron stars~\cite{Roberts}. Moreover, it is interesting
to note that the narrow uncertain range for the proposed Einstein
Telescope will enable it to tightly constrain the symmetry energy
especially at supra-saturation densities.

\section{Summary}
In summary, significant progress has been made in recent years in
constraining the symmetry energy mostly around and below the
saturation density using both terrestrial nuclear laboratory data
and astrophysical observations. However, to fully understand the
nature of neutron-rich nucleonic matter and its equation of state
especially at supra-saturation densities many interesting questions
remain to be studied. In particular, the high-density behavior of
the symmetry energy is still among the most uncertain properties of neutron-rich matter.
Besides continuing the search of sensitive observables in terrestrial experiments and
astrophysical observations, it is important to understand the
underlying physics leading to the divergent predictions for the
high-density symmetry energy. In particular, effects of the
tensor-force induced short-range neutron-proton correlation and
the short-range behavior of the tensor force itself on the
high-density behavior of the symmetry energy deserve some special
attention. Given the strong ongoing efforts and close collaborations of many people in
both the nuclear physics and astrophysics communities, it is expected that 
more stringent constraints on the symmetry energy over a broad density range will come soon.

\section{Acknowledgments}
This work is supported in part by the US National Aeronautics and Space
Administration under grant NNX11AC41G issued through the Science
Mission Directorate, the US National Science Foundation under
Grants No. PHY-0757839, No. PHY-1062613 and No. PHY-1068022, the NNSF of China under Grant Nos. 10975097, 11135011, 11275125 and 11175085,
the Shanghai Rising-Star Program under grant No. 11QH1401100, the ``Shu Guang"
project supported by Shanghai Municipal Education Commission and Shanghai
Education Development Foundation, the Program for Professor of Special
Appointment (Eastern Scholar) at Shanghai Institutions of Higher Learning,
and the Science and Technology Commission of Shanghai Municipality (11DZ2260700).

\end{document}